\documentclass[bibyear]{aa} 
\usepackage{epsfig}
\usepackage{amsmath}
\usepackage{subfigure}
\usepackage{natbib}
\usepackage{multirow}
\usepackage{color}
\usepackage{longtable}
\usepackage{graphicx}
\usepackage{epstopdf}
\usepackage{booktabs}
\usepackage{txfonts}
\usepackage{graphicx}
\usepackage{txfonts}
\usepackage{color}
\usepackage{stfloats}
\usepackage{amsmath}
\usepackage{multirow}
\usepackage{epsfig}
\usepackage{amsmath}
\usepackage{subfigure}
\usepackage{natbib}
\usepackage{longtable}
\usepackage{graphicx}
\usepackage{epstopdf}
\usepackage{booktabs}
\usepackage{txfonts}
\usepackage[utf8]{inputenc}%
\usepackage{graphicx}
\usepackage{txfonts}
\usepackage{stfloats}

\newcommand{\tablenotea}[1]{\parbox{9.0cm}{\indent \footnotesize{#1}}}

\newcommand{\kms}{km\,s$^{-1}$}

\newcommand{\jms}{J. Mol. Spectr.}

\newcommand{\prev}{Phys. Rev.}

\begin{document}

\title{Discovery of the elusive thioketenylium, HCCS$^+$, in TMC-1\thanks{Based on
observations carried out with the Yebes 40m telescope (projects 19A003, 20A014, 20D023, and 21A011) and the Institut de Radioastronomie Millim\'etrique (IRAM) 30m telescope. The 40m radio telescope at Yebes Observatory is  operated by the Spanish Geographic Institute
(IGN, Ministerio de Transportes, Movilidad y Agenda Urbana). IRAM is supported by INSU/CNRS (France), MPG (Germany), and IGN (Spain).}}

\author{
C.~Cabezas\inst{1},
M.~Ag\'undez\inst{1},
N.~Marcelino\inst{2,3},
B.~Tercero\inst{2,3},
Y.~Endo\inst{4},
R.~Fuentetaja\inst{1},
J.~R.~Pardo\inst{1},
P.~de~Vicente\inst{3},
and
J.~Cernicharo\inst{1}
}

\institute{Grupo de Astrof\'isica Molecular, Instituto de F\'isica Fundamental (IFF-CSIC), C/ Serrano 121, 28006 Madrid, Spain.
\email carlos.cabezas@csic.es; jose.cernicharo@csic.es
\and Observatorio Astron\'omico Nacional (IGN), C/ Alfonso XII, 3, 28014, Madrid, Spain.
\and Centro de Desarrollos Tecnol\'ogicos, Observatorio de Yebes (IGN), 19141 Yebes, Guadalajara, Spain.
\and Department of Applied Chemistry, Science Building II, National Yang Ming Chiao Tung University, 1001 Ta-Hsueh Rd., Hsinchu 300098, Taiwan
}

\date{Received; accepted}

\abstract{We report the detection in TMC-1 of the cation HCCS$^+$ ($\tilde{X}$ $^3\Sigma^-$), which
is the protonated form of the widespread radical CCS. This is the first time that a protonated radical
has been detected in a cold dark cloud. Twenty-six hyperfine components from twelve rotational transitions
have been observed with the Yebes 40m and IRAM 30m radio telescopes. We confidently assign the
characteristic rotational spectrum pattern to HCCS$^+$ based on the good agreement between the
astronomical and theoretical spectroscopic parameters. The column density of HCCS$^+$ is
(1.1$\pm$0.1)$\times$10$^{12}$ cm$^{-2}$, and the CCS/HCCS$^+$ abundance ratio is 50$\pm$10,
which is very similar to that of CS/HCS$^+$ (35$\pm$8) and CCCS/HCCCS$^+$ (65$\pm$20). From a
state-of-the-art gas-phase chemical model, we conclude that HCCS$^+$ is mostly formed by
reactions of proton transfer from abundant cations such as HCO$^+$, H$_3$O$^+$, and H$_3^+$
to the radical CCS.}


\titlerunning{Interstellar HCCS$^+$ radical}
\authorrunning{Cabezas et al.}

\maketitle

\section{Introduction}

The cold dark cloud TMC-1 shows a rich and complex chemistry, and it is known to
contain a variety of protonated species of abundant neutral molecules. In addition to
the widespread ions HCO$^+$ and N$_2$H$^+$, the polyatomic cations detected in TMC-1 are
HCS$^+$ \citep{Irvine1983}, HCNH$^+$ \citep{Schilke1991}, HC$_3$NH$^+$ \citep{Kawaguchi1994},
HCO$_2^+$ {\citep{Turner1999}, NCCNH$^+$ \citep{Agundez2015}, HC$_5$NH$^+$
\citep{Marcelino2020}, HC$_3$O$^+$ \citep{Cernicharo2020}, HC$_3$S$^+$ \citep{Cernicharo2021a},
and CH$_3$CO$^+$ \citep{Cernicharo2021b}. The abundance ratio between a protonated molecule and
its neutral counterpart, MH$^+$/M, is sensitive to the degree of ionization and to the formation
and destruction rates of the cation \citep{Agundez2015}. The main formation route to the cation is
usually the proton transfer to the neutral, while its main destruction process is the dissociative
recombination with electrons. It is interesting to note that both chemical models and observations
suggest a trend in which the abundance ratio MH$^+$/M increases with increasing proton affinity
of M \citep{Agundez2015}. Thus, protonated species of abundant molecules with high proton affinities
are good candidates for detection.

The number of sulfur-bearing species detected to date in TMC-1 is small compared to oxygen- and nitrogen-bearing species (see e.g. \citealt{McGuire2019}). However, the most recent discoveries have revealed that TMC-1 is also a source rich in sulfur-bearing molecules \citep{Cernicharo2021a, Cernicharo2021c,Cernicharo2021d}. The carbon chains CCS and CCCS are particularly abundant in this cloud
\citep{Saito1987,Yamamoto1987}. The protonated form of the CCCS molecule, HC$_3$S$^+$, was discovered in TMC-1 and confirmed in the laboratory by \citet{Cernicharo2021a}. The protonated form of CCS, HCCS$^+$, has not been detected in space so far, but it is considered a promising candidate to be observed \citep{Agundez2015} given that CCS is even more abundant than CCCS, CCS/CCCS $\sim$4.2, it has a high calculated proton affinity, between 869.6 \citep{Maclagan1992} and 901.2 kJ mol$^{-1}$, \citep{Barrientos1991}, and chemical models predict HCCS$^+$ to be just a few times less abundant than CCS \citep{Agundez2015}. However, the lack of laboratory data for
HCCS$^+$ has hampered its detection.

\begin{figure*}[]
\centering
\includegraphics[angle=0,width=0.95\textwidth]{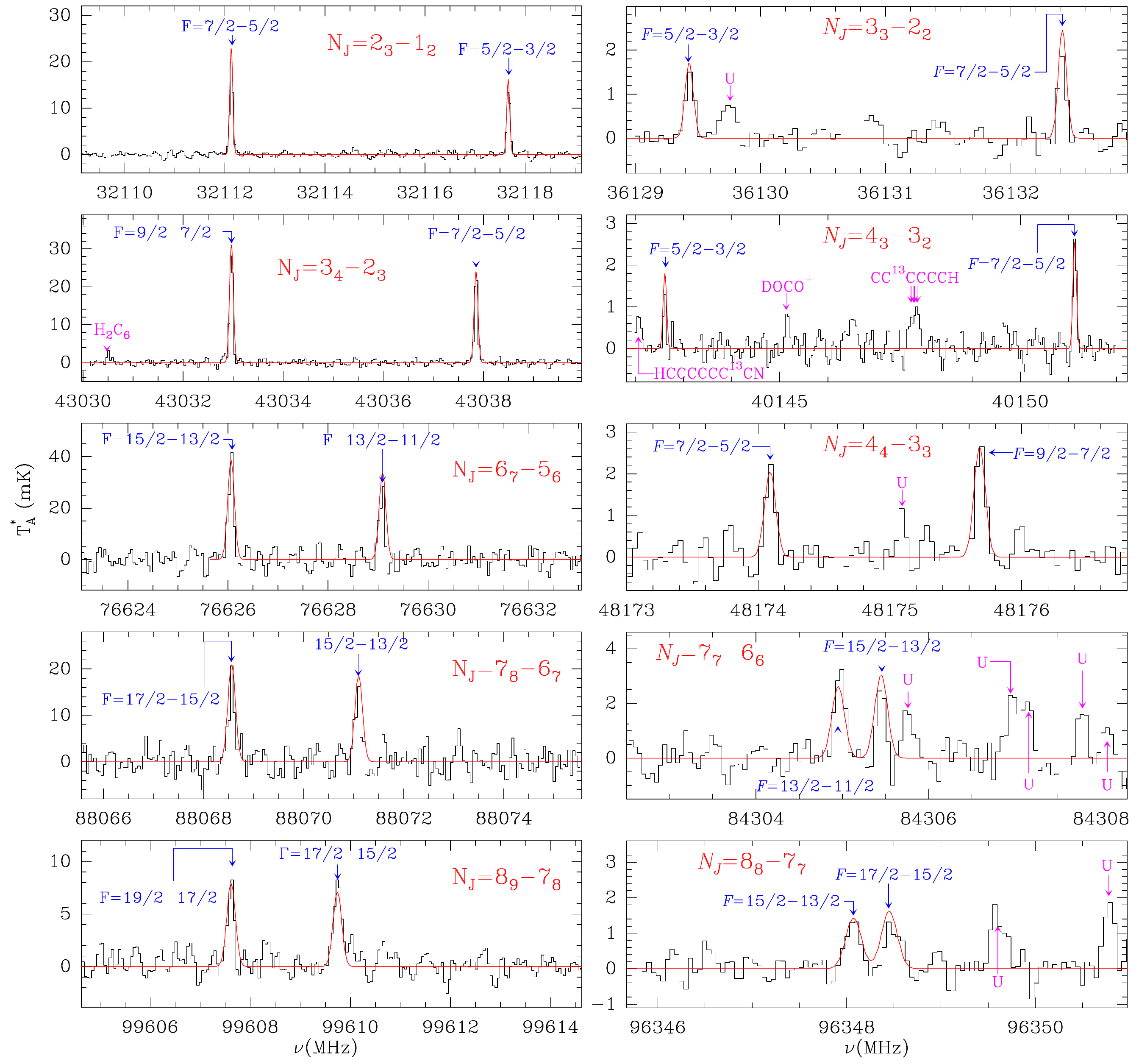}
\caption{Spectra of the observed transitions of HCCS$^+$ in TMC-1. The abscissa corresponds
to the rest frequency assuming a local standard of rest velocity of 5.83\,\kms. The ordinate
is antenna temperature in millikelvin. Blanked channels correspond to negative features produced
in the folding of the frequency-switching data. The red lines correspond to the synthetic spectra
calculated for a column density of 1.1$\times$10$^{12}$ cm$^{-2}$ and a rotational temperature of
7.0\,K for $N_u\le$4 and 4.5\,K for higher $N_u$ values (see Sect. \ref{excitation}).}
\label{fig_hccs+}
\end{figure*}

High resolution line surveys can be used to conduct molecular spectroscopy in space; they are an invaluable
alternative that can be used to overcome the lack of laboratory experimental data. Some examples of these kinds of
identifications carried out by our group are C$_5$H \citep{Cernicharo1986}, C$_5$N$^-$
\citep{Cernicharo2008}, MgC$_3$N and MgC$_4$H \citep{Cernicharo2019}, HC$_5$NH$^+$ \citep{Marcelino2020},
MgC$_5$N and MgC$_6$H \citep{Pardo2021}, and H$_2$NC \citep{Cabezas2021a}. If the source displays narrow
lines and spectral resolution is sufficiently high, it is possible to resolve the hyperfine structure and derive
very precise spectroscopic parameters \citep{Cabezas2021a}. Here we present a new case of molecular
spectroscopy done in space. Thanks to the sensitivity of our QUIJOTE\footnote{\textbf{Q}-band
\textbf{U}ltrasensitive \textbf{I}nspection \textbf{J}ourney to the \textbf{O}bscure \textbf{T}MC-1
\textbf{E}nvironment} line survey of TMC-1 \citep{Cernicharo2021e} we have detected various groups
of unidentified lines, revealing a characteristic fine and hyperfine structure, which we confidently
assign to HCCS$^+$ based on our quantum chemical calculations.

\section{Observations}

The observations in the Q band are part of the QUIJOTE spectral line survey, carried out towards TMC-1 ($\alpha_{J2000}=4^{\rm h} 41^{\rm  m} 41.9^{\rm s}$ and $\delta_{J2000}=+25^\circ 41' 27.0''$) with the Yebes 40m radio telescope. The QUIJOTE data currently available were taken during various observing sessions between November 2019 and April 2021. These observations were carried out using the frequency-switching technique with a frequency throw of 10\,MHz  during the first observing runs and of 8\,MHz in the later ones. Different frequency coverages were observed, 31.08-49.52 GHz and 31.98-50.42 GHz, which permitted us to verify that no spurious ghosts were produced in the down-conversion chain. In this process, the signal coming from the receiver is down-converted to 1-19.5 GHz and then split into eight bands, each with a coverage of 2.5 GHz, which are analysed by the fast Fourier transform (FFT) spectrometers.

Data in the 3\,mm band come from a spectral line survey performed towards TMC-1 and B1-b using the IRAM 30m telescope \citep{Marcelino2007,Cernicharo2012}. Additional data were gathered in September and October 2021. For these observations the 3mm EMIR receiver was used connected to a FFT spectrometer, providing a spectral resolution of 48.84 kHz. We observed two setups at slightly different central frequencies in order to check for spurious signals, line emission from the image band, and other technical artefacts. The observations were performed in the frequency-switching observing mode with a frequency throw of 18 MHz. System temperatures varied between 80 and 130 K, and the final rms for the line studied in this work is between 1.1 and 3 mK.

The intensity scale used for the Yebes 40m and IRAM 30m observations, the antenna temperature $T_A^*$, was calibrated using two  absorbers at different temperatures and the atmospheric transmission model ATM \citep{Cernicharo1985, Pardo2001}. Calibration  uncertainties were adopted to be 10~\%. All data were analysed using the GILDAS package\footnote{\texttt{http://www.iram.fr/IRAMFR/GILDAS}}.

\section{Results and discussion}

\subsection{Discovery of HCCS$^+$}

The ion HCCS$^+$ is predicted to be an abundant interstellar species \citep{Agundez2015}. This open-shell cation was investigated by \citet{Puzzarini2008} using high level quantum chemical calculations. In this work, the equilibrium geometry and energy for HCCS$^+$ are presented at different levels of theory. Using the best estimated equilibrium structure for HCCS$^+$ in its $^3\Sigma^-$ ground electronic state, we calculate a $B$ rotational constant of 6048 MHz. This value has been used in the past to make frequency predictions
and search for this species towards TMC-1 \citep{Cernicharo2021a}, with a 3$\sigma$ upper limit to its column density of 9$\times$10$^{11}$ cm$^{-2}$ obtained. However, due to the $^3\Sigma$ ground state for HCCS$^+$, its rotational spectral pattern is not as simple as that of other closed-shell species. The prediction of this spectral pattern requires knowing not only the rotational constant, but also additional spectroscopic parameters, such as the spin-spin interaction constant, $\lambda$, the spin-rotation interaction constant, $\gamma$, and the hydrogen nuclei hyperfine constants $b_F$ and $c$. For our predictions we calculated all these molecular parameters (see Sect. \ref{spectroscopy}) except $\lambda$, for which we adopted the value derived for the CCS ($^3\Sigma^-$) radical \citep{Yamamoto1990}, which is isoelectronic to HCCS$^+$. These predictions have very large uncertainties, and the search was unsuccessful even for the $J$=$N$ lines, which could appear at harmonically related frequencies. Moreover, the expected
intensity variation in the lines was also an important unknown, and a handicap, because the $J=N+1$ levels are significantly lower in energy than those with $J=N$ and $J=N-1$ (see Sect. \ref{excitation}). Hence, the three rotational lines arising from a given $N$ could have very different intensities in a cold dark cloud such as TMC-1.

In a new attempt, we followed a different strategy to find lines that could be attributable to HCCS$^+$. The expected pattern for the most intense lines ($J=N+1$) consists in doublets of lines with a separation of a few megahertz. We searched for these doublets in a very wide region and found a couple of very strong unidentified lines (the strongest among our U-lines) at 32.11 GHz, which is 350 MHz lower to the initially predicted frequency. A second pair of lines with similar intensity was found at 43.03 GHz. For a $^3\Sigma$ molecule the transitions $J$=$N$ follow a very good harmonic relation with a very small error. However, the energy terms for the levels $J=N-1$ and $J=N+1$ contain contributions from $\lambda$ and $\gamma$, and the frequency ratio between subsequent transitions deviates from the ratio of the $J$ values. The frequency ratio between the two observed doublets is near 4/3, with an error of a few per cent, which is in agreement with the expected ratio for the $J=N+1$ transitions. Hence, these two doublets, shown in Fig. \ref{fig_hccs+}, can be assigned to the $N_J=2_3-1_2$ and $3_4-2_3$ transitions of HCCS$^+$. Using this near-harmonic relation, we predicted the frequencies of the next doublets of lines within our data at 3\,mm and found three additional doublets that can be assigned to the $N_J=6_7-5_6$, $7_8-6_7$, and $8_9-7_8$ transitions. As expected, the separation between the two components of each doublet decreases with $N$ (see Fig. \ref{fig_hccs+}). These five pairs of lines can be fitted with a standard Hamiltonian for a $^3\Sigma$ molecule (see Sect. \ref{spectroscopy}). New frequency predictions, improved by the assigned transitions, were used to search for the $J=N$ and $J=N-1$ transitions. Three of them, the $N_J=3_3-2_2, 4_3-3_2$, and $4_4-3_3$, were easily found in the QUIJOTE line survey (see Fig. \ref{fig_hccs+}). Once merged with the previous lines, the predictions for additional lines at 3\,mm permitted four additional doublets corresponding to the $N_J=7_7-6_6$, $8_8-7_7$, $7_6-6_5$, and $8_7-7_6$ transitions to be detected. The final set of twenty-six hyperfine components includes two weak components from the $N_J=2_3-1_2$ and $3_4-2_3$ transitions. The derived line parameters are given in Table \ref{table:fits}. The energy levels associated with the observed rotational transitions are shown in Fig. \ref{fig_ener} (see Sect. \ref{excitation}).

\subsection{Rotational analysis of HCCS$^+$}
\label{spectroscopy}

\begin{table}
\small
\caption{Spectroscopic parameters of HCCS$^+$, in MHz.}
\label{table:constants}
\centering
\begin{tabular}{lccc}
\hline
\hline
\multicolumn{1}{c}{Parameter}  & \multicolumn{1}{c}{TMC-1 fit} & \multicolumn{1}{c}{Theoretical} & \multicolumn{1}{c}{CCS\,$^a$}\\
\hline
$B$                       &  ~  6021.89878(55)\,$^b$      &  ~ 6021.2\,$^{c,d}$    &  ~ 6477.75036(71)      \\
$D$                       &  ~   0.0012543(72)            &  ~ 0.00120\,$^{e}$     &  ~ 0.00172796(95)      \\
$\lambda$                 &  ~   108970.78(83)            &  ~   -                 &  ~ 97196.07(77)        \\
$\lambda_D$               &  ~    0.04060(65)             &  ~   -                 &  ~ 0.02700(67)         \\
$\gamma$                  &  ~   $-$41.776(46)            &  ~   $-$18.4\,$^{e}$   &  ~ $-$14.737(49)       \\
$b_F$$^{\rm(H)}$          &  ~   $-$44.961(23)            &  ~   $-$47.6\,$^{f}$   &  ~      -              \\
$c$$^{\rm(H)}$            &  ~    31.663(70)              &  ~    71.6\,$^{f}$     &  ~      -              \\
$\sigma$\,$^{g}$                  &  ~        26.1                &  ~    -                &  ~      18             \\
$N$                       &  ~        26                  &  ~    -                &  ~      31             \\
\hline
\hline
\end{tabular}
\tablenotea{$^a$\,Values from \citet{Yamamoto1990}. $^b$\,Numbers in parentheses are 1$\sigma$ uncertainties in units of the last digits. $^c$\,$B$ value was corrected using the experimental and theoretical constants of CCS (see text). $^d$\,RCCSD(T)-F12/cc-pCVTZ-F12.  $^e$\, UMP2/cc-pVTZ. $^f$\,QCISD/cc-pVTZ. $^g$\,In units of kHz.}
\end{table}

All the observed hyperfine components were analysed with the SPFIT program \citep{Pickett1991} using a Hamiltonian for a linear molecule in a $^3\Sigma$ electronic state. Since HCCS$^+$ contains one nucleus with a non-zero nuclear spin, the appropriate Hamiltonian to describe the rotational levels can be written as follows:
\begin{equation}
H = H_{rot} + H_{ss} + H_{sr} + H_{mhf}
,\end{equation}
where $H_{rot}$, $H_{ss}$, and $H_{sr}$ denote the rotational, spin-spin, and spin-rotation terms, respectively, and $H_{mhf}$ represents the magnetic hyperfine coupling interaction term due to the hydrogen nucleus. The coupling scheme used is \textbf{J}\,=\,\textbf{N}\,+\,\textbf{S}, \textbf{F}\,=\,\textbf{J}\,+\,\textbf{I}(H).

The molecular constants determined from the fit are given in Table \ref{table:constants} together with those predicted by ab initio calculations \citep{Frisch2016,Werner2020} and those for the CCS radical. All the constants are very well determined, indicating that the fine and hyperfine structures are well described by the employed Hamiltonian. The identification of HCCS$^+$ as the carrier of the observed lines is unequivocal. The rotational constant $B$ obtained by us through ab initio calculations (RCCSD(T)-F12/cc-pCVTZ-F12; \citealt{Knizia2009,Hill2010}) is 6042.5 MHz, which is similar to that calculated by \cite{Puzzarini2008} and deviates from the value derived from \mbox{TMC-1} data by just 0.3\,\%. This difference decreases down to 0.01\,\% when the theoretical value is scaled using the experimental/theoretical ratio of the isoelectronic CCS radical. The $B$ value obtained for CCS at the same level of theory is 6500.7 MHz. The value derived for $D$ is also in good agreement with that predicted at the MP2/cc-pVTZ level of theory \citep{Moller1934,Woon1993}. The values for the spin-spin interaction constant and its centrifugal distortion constant, $\lambda$ and $\lambda_D$, cannot be estimated using quantum chemical calculations. However, their values are similar to those found for the isoelectronic CCS radical, as expected. The value for the spin-rotation constant, $\gamma$, is significantly different from that predicted by ab initio calculations \citep{Moller1934,Woon1993}. For the CCS radical, the predicted value for $\gamma$ at the same level of theory is $-$110.2 MHz, quite different from the experimental value as well. The accuracy of the prediction of $\gamma$ is much smaller than that of other molecular constants. For the hyperfine constants, the agreement with the theoretical values is disparate. The derived value for the Fermi contact constant, $b_F$, agrees well with the predicted value (QCISD/cc-pVTZ; \citealt{Pople1987,Woon1993}). However, for the dipole-dipole constant, $c$, the discrepancy is very large. This is very surprising because the prediction of hyperfine constants at the employed level of theory usually gives satisfactory results. In order to get some insight into this, we calculated the $c$ constant for the other $^3\Sigma^-$ molecule, HCCN, and we found the same behaviour. The predicted value, at the same level of theory, for the $c$ constant is 52.8 MHz, while the experimental one is 32.91(18) MHz \citep{Endo1993}. The prediction of the $c$ constant for $^3\Sigma^-$ molecules may not be as accurate as expected.

\subsection{The abundance and excitation of HCCS$^+$ in \mbox{TMC-1}}
\label{excitation}

\begin{figure}[]
\centering
\includegraphics[angle=0,width=0.43\textwidth]{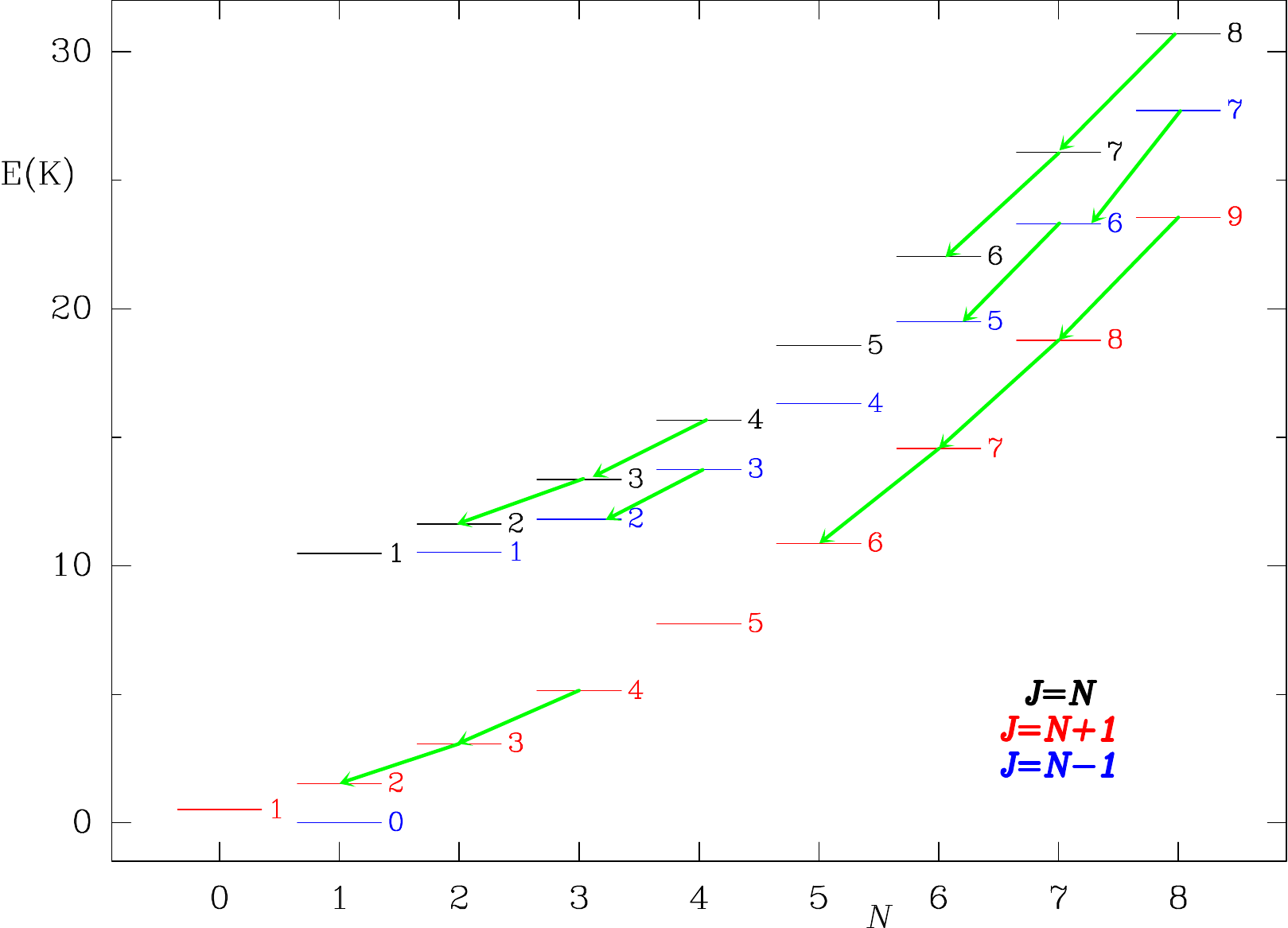}
\caption{Energy levels of HCCS$^+$ organized by $N$ and $J$ values.
The observed transitions are indicated by green arrows.}
\label{fig_ener}
\end{figure}

The scheme of rotational energy levels of a $^3\Sigma$ molecule has a peculiar pattern, as shown in Fig. \ref{fig_ener} for HCCS$^+$. Since the $\lambda$ value is rather large, similar to that of CCS, the $J=N+1$ ladder of levels is always below the $J=N-1$ and $J=N$ ladders, the latter being the highest in energy. For a cloud with a kinetic temperature of 10\,K, the energy difference between these three ladders affects the intensity of the rotational transitions. As shown in Fig. \ref{fig_hccs+}, the $J=N+1$ lines are very strong, but their intensity declines for high $N$ values. The upper level energy of the highest $N$ transition in our data, $8_9-7_8$, is $\sim$24\,K. If we assume thermalization at 10 K and optically thin emission, this line should be as strong as the $3_4-2_3$ one, but this is not the case. Hence, the rotational levels of HCCS$^+$ are not thermalized. This is in fact indicated by a rotation diagram constructed with the lines observed, where two different rotational temperatures are seen. The lines observed with QUIJOTE are reproduced with a rotational temperature of 7.0$\pm$1.5\,K and a column density of (1.1$\pm$0.2)$\times$10$^{12}$ cm$^{-2}$, while the lines observed at 3\,mm, which correspond to transitions involving higher level energies, are reproduced with the same column density but a rotational temperature of 4.5$\pm$1.0\,K.

In order to evaluate possible excitation effects in the different lines of HCCS$^+$, we used the MADEX code to perform a large velocity gradient calculation (MADEX follows the formalism described by \citealt{Goldreich1974}). The hyperfine structure was neglected in these calculations. For this purpose, we adopted, as done by \citet{Cernicharo2021f} for CCO, the collisional rates of OCS \citep{Green1978} and the infinite-order sudden approximation for a $^3\Sigma$ molecule \citep{Alexander1983,Corey1983,Fuente1990}. For a volume density of H$_2$ of 4$\times$10$^4$ cm$^{-3}$ \citep{Cernicharo1987}, the excitation temperatures obtained using this approach are $\sim$7-10\,K for the lines observed in the Q band. However, the excitation temperature decreases to 5.7\,K for $N_u$=6 and to 4.5\,K for $N_u$=8. We explored the effect of n(H$_2$) on the excitation temperatures of HCCS$^+$. For n(H$_2$)$\ge10^4$ cm$^{-3}$, the $N$=2 and 3 ($J=N+1$) lines have an excitation temperature close to T$_K$. However, the lines associated with upper levels with $N\ge$ 6 show a sharp dependence of T$_{exc}$ with density. The best fit to the observations is obtained for n(H$_2$)=2-4$\times$10$^4$ cm$^{-3}$ and N(HCCS$^+$)=(1.1$\pm$0.2)$\times$10$^{12}$ cm$^{-2}$. We adopted T$_{rot}$=7.0\,K for $N_u\le$4 and 4.5\,K for higher $N_u$ values and a source radius of 40$''$ \citep{Fosse2001}. The match between the observations is rather good despite the uncertainty in the collisional rates (see Fig.~\ref{fig_hccs+}). The derived column density is only 20\% larger than the previous 3$\sigma$ limit reported by \citet{Cernicharo2021a}.

\subsection{The chemistry of HCCS$^+$}
\label{sec:chemistry}

\begin{figure}
\centering
\includegraphics[angle=0,width=0.95\columnwidth]{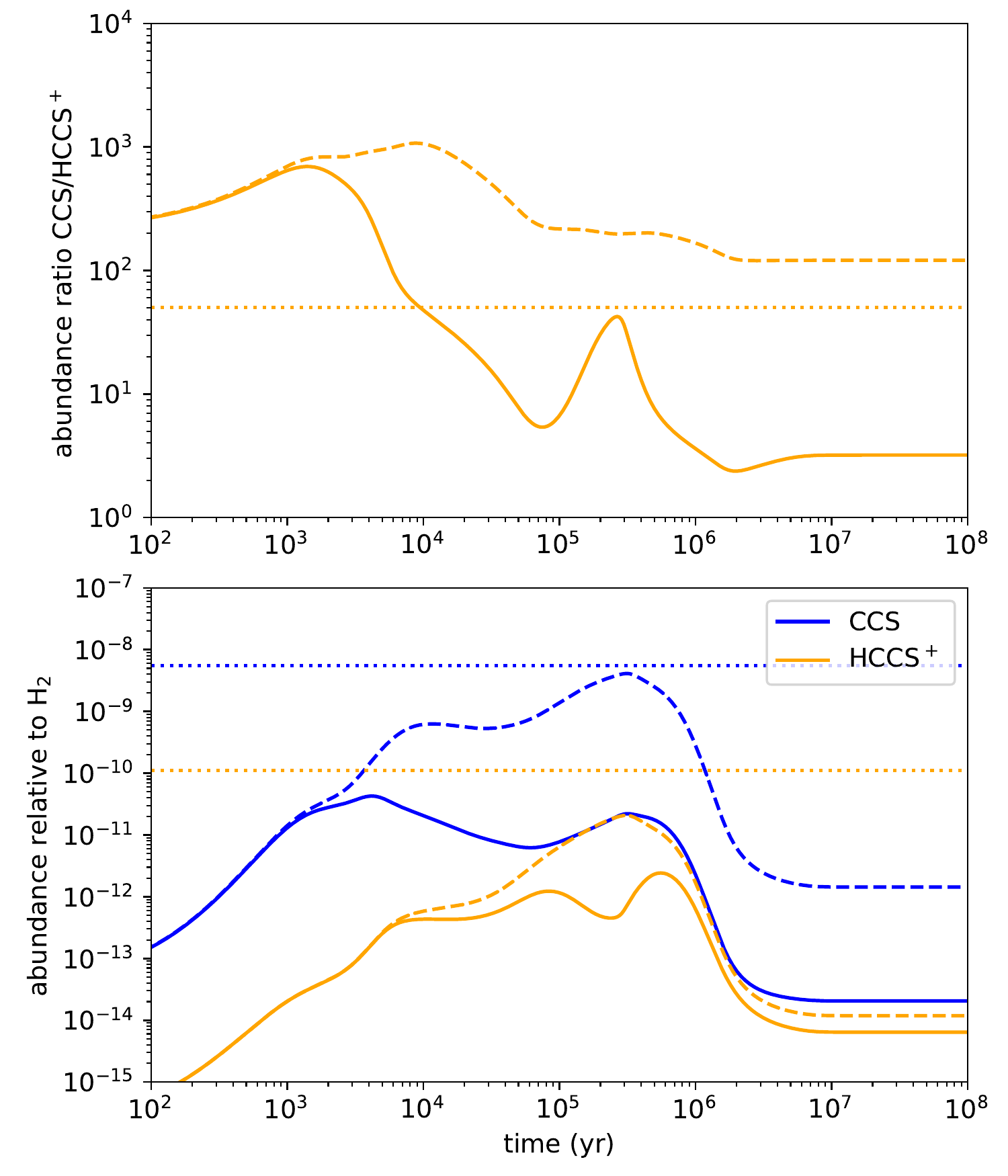}
\caption{Calculated abundances of CCS and HCCS$^+$ (\textit{bottom}) and abundance ratio
CCS/HCCS$^+$ (\textit{top}) as a function of time. Dashed lines correspond to a chemical model in which the reaction O + CCS is removed from the chemical network. Observed values are indicated by horizontal dotted lines. }
\label{fig:abun}
\end{figure}

Adopting the column density of CCS derived by \citet{Cernicharo2021a} of (5.5$\pm$0.6)$\times$10$^{13}$ cm$^{-2}$, the abundance ratio between CCS and its protonated form is 50$\pm$10. This ratio is very similar to that obtained by those authors for CS and HCS$^+$ (35$\pm$8) and for CCCS and HCCCS$^+$ (65$\pm$20). In order to get insight into the chemistry of HCCS$^+$ in \mbox{TMC-1,} we constructed a pseudo-time-dependent gas-phase chemical model with parameters typical of cold dark clouds (see \citealt{Cernicharo2021a} for details and \citealt{Agundez2015} for further details on the adopted physical conditions). We adopted the chemical network {\small \texttt{kida.uva.2014}} from the KIDA database \citep{Wakelam2015}.

The calculated abundances of CCS and HCCS$^+$ are shown in the bottom panel of Fig.~\ref{fig:abun}, with the calculated abundance ratio CCS/HCCS$^+$ shown in the top panel of the same figure. If we first focus on the neutral-to-protonated ratio, we see that the calculated value at steady state is pretty small, $\sim$\,3 (solid orange curve in the top panel of Fig.~\ref{fig:abun}), in agreement with the prediction by \cite{Agundez2015}. However, steady state may not be a good choice for \mbox{TMC-1}. Instead, the so-called early time (10$^5$-10$^6$ yr) may better represent the state of \mbox{TMC-1} since at this time calculated and observed abundances show the best overall agreement (see e.g. \citealt{Agundez2013}). At a time of (2-3)\,$\times$\,10$^5$ yr, the calculated CCS/HCCS$^+$ ratio agrees very well with the observed value. However, the fractional abundance of HCCS$^+$ is below the observed value by about two orders of magnitude. According to the chemical model, HCCS$^+$ is mostly destroyed through dissociative recombination with electrons, as usual for cations. On the other hand, the formation of HCCS$^+$ is mostly driven by ion-neutral reactions, such as S$^+$ + C$_2$H$_2$, S + C$_2$H$_3^+$, and S + C$_2$H$_4^+$, rather than by reactions of proton transfer to CCS from HCO$^+$, H$_3$O$^+$, and H$_3^+$. This occurs because the chemical model severely underestimates the abundance of CCS, by two to three orders of magnitude (see the solid blue curve in the bottom panel of Fig.~\ref{fig:abun}). This has been already noticed by \cite{Cernicharo2021a} and points to an inadequate description of the chemistry of CCS.

The {\small \texttt{kida.uva.2014}} network includes a very efficient process of destruction of CCS through reaction with O atoms. The rate coefficient for this reaction has been estimated by \cite{Loison2012} to be as high as 10$^{-10}$ cm$^3$ s$^{-1}$. If we remove this reaction from the chemical network, the calculated abundance of CCS at early time agrees much better with the observed value (dashed blue curve in the bottom panel of Fig.~\ref{fig:abun}), and the abundance of protonated CCS is also enhanced because the reactions of proton transfer to CCS become the dominant routes to HCCS$^+$. In this case, the calculated abundance of HCCS$^+$ at early time is closer to the observed value, while the neutral-to-protonated ratio is still of the order of the value observed. It would therefore be interesting to study the reaction O + CCS and evaluate how rapid it could be at low temperatures. Whatever the failure in the description of the chemistry of CCS, it seems clear that proton transfer to CCS should be the main pathway to HCCS$^+$. In such a case, the neutral-to-protonated abundance ratio can be evaluated through the expression \citep{Agundez2015}
\begin{equation}
{\rm CCS/HCCS^+} = \frac{k_{DR}}{k_{PT}} \rm{\frac{[e^-]}{[XH^+]}}, \label{eq:ratio}
\end{equation}
where $k_{DR}$ and $k_{PT}$ are the rate coefficients for dissociative recombination of HCCS$^+$ and proton transfer to CCS, respectively, while [e$^-$] and [XH$^+$] are the abundances of electrons and proton donors, respectively. Assuming to a first order that [XH$^+$] = [e$^-$], and adopting $k_{DR}$(10\,K)\,=\,1.6\,$\times$10$^{-6}$ cm$^3$ s$^{-1}$ and $k_{PT}$(10\,K)\,=\,1.6\,$\times$10$^{-8}$ cm$^3$ s$^{-1}$, as given in the {\small \texttt{kida.uva.2014}} network, the predicted CCS/HCCS$^+$ ratio is 100, which is only a factor of two higher than the value derived from the observations. This supports the idea that HCCS$^+$ is mainly formed by proton transfer to CCS from the most abundant proton donors.

\section{Conclusions}

We observed a total of 26 unidentified lines in the cold dark cloud TMC-1, which we confidently assign to the cation HCCS$^+$ based on high level ab initio calculations. The astronomical lines are used to precisely characterize the rotational spectrum of HCCS$^+$ for the first time. The analysis of the line intensities considering possible excitation effects provides a column density for HCCS$^+$ of (1.1$\pm$0.2)$\times$10$^{12}$ cm$^{-2}$ with rotational temperatures of 7.0$\pm$1.5\,K and 4.5$\pm$1.0\,K for the transitions observed in the Q band and at 3\,mm, respectively. According to our chemical modelling calculations, in {TMC-1} the cation HCCS$^+$ is most likely formed through proton transfer to CCS from the most abundant proton donors, such as HCO$^+$, H$_3$O$^+$, and H$_3^+$.

\begin{acknowledgements}

This research has been funded by ERC through grant ERC-2013-Syg-610256-NANOCOSMOS. Authors also thank Ministerio de Ciencia e Innovaci\'on for funding support through projects PID2019-106235GB-I00 and PID2019-107115GB-C21 / AEI / 10.13039/501100011033. MA thanks Ministerio de Ciencia e Innovaci\'on for grant RyC-2014-16277.

\end{acknowledgements}

\begin{appendix}

\section{Transition frequencies and line parameters for HCCS$^+$ towards TMC-1} \label{appen}

Line parameters were derived from a Gaussian fit to the observed lines using \texttt{GILDAS}. Observed frequencies were derived assuming a local standard of rest velocity of 5.83 km\,s$^{-1}$.

\begin{table*}
\begin{center}
\caption{Observed lines of HCCS$^+$ towards TMC-1.}
\label{table:fits}
\begin{tabular}{cllrrlrc}
\hline\hline
\multicolumn{1}{c}{Transition}  & \multicolumn{1}{c}{$\nu_{obs}$\,$^a$} & \multicolumn{1}{c}{$\nu_{cal}$\,$^b$} & \multicolumn{1}{c}{$\nu_{obs}$-$\nu_{cal}$} &
\multicolumn{1}{c}{$\int T_{\rm A}^* dv$\,$^c$} & \multicolumn{1}{c}{$\Delta$v\,$^d$} & \multicolumn{1}{c}{$T_{\rm A}^*$\,$^e$} & Notes \\
$(N_J,F)_{\rm u}-(N_J,F)_{\rm l}$ & \multicolumn{1}{c}{(MHz)} & \multicolumn{1}{c}{(MHz)} & \multicolumn{1}{c}{(MHz)} &
\multicolumn{1}{c}{(mK\,km\,s$^{-1}$)} & \multicolumn{1}{c}{(km\,s$^{-1}$)} & \multicolumn{1}{c}{(mK)}   &   \\
\hline
$2_3,7/2-1_2,5/2$  &  32112.130$\pm$0.010  &  32112.1277$\pm$0.0044  &  0.0023  & 15.38$\pm$0.14 & 0.76$\pm$0.01 & 19.05$\pm$0.16 & \\
$2_3,5/2-1_2,3/2$  &  32117.667$\pm$0.010  &  32117.6619$\pm$0.0035  &  0.0051  & 10.70$\pm$0.17 & 0.75$\pm$0.01 & 13.40$\pm$0.18 & \\
$2_3,5/2-1_2,5/2$  &  32133.510$\pm$0.020  &  32133.5152$\pm$0.0053  & -0.0052  &  0.72$\pm$0.06 & 0.80$\pm$0.08 &  0.84$\pm$0.12 & \\ 

$3_3,7/2-2_2,5/2$  &  36129.436$\pm$0.010  &  36129.4310$\pm$0.0056  &  0.0050  &  0.96$\pm$0.10 & 0.64$\pm$0.07 &  1.42$\pm$0.18 & \\
$3_3,5/2-2_2,3/2$  &  36132.410$\pm$0.010  &  36132.4146$\pm$0.0047  & -0.0046  &  0.99$\pm$0.13 &    0.60$^f$   &  1.54$\pm$0.18 & A \\ 

$4_3,5/2-3_2,3/2$  &  40142.608$\pm$0.020  &  40142.6102$\pm$0.0069  & -0.0022  &  0.52$\pm$0.08 & 0.28$\pm$0.38 &  1.71$\pm$0.18 & \\ 
$4_3,7/2-3_2,5/2$  &  40151.134$\pm$0.010  &  40151.1287$\pm$0.0069  &  0.0053  &  1.73$\pm$0.12 & 0.59$\pm$0.04 &  2.75$\pm$0.23 & \\

$3_4,9/2-2_3,7/2$  &  43032.965$\pm$0.010  &  43032.9644$\pm$0.0045  &  0.0006  & 17.33$\pm$0.12 & 0.56$\pm$0.01 & 28.89$\pm$0.23 & \\
$3_4,7/2-2_3,5/2$  &  43037.857$\pm$0.010  &  43037.8520$\pm$0.0037  &  0.0050  & 13.64$\pm$0.12 & 0.56$\pm$0.01 & 22.82$\pm$0.23 & \\
$3_4,7/2-2_3,7/2$  &  43059.244$\pm$0.015  &  43059.2395$\pm$0.0067  &  0.0045  &  0.80$\pm$0.19 & 0.86$\pm$0.21 &  0.88$\pm$0.20 & \\ 

$4_4,7/2-3_3,5/2$  &  48174.092$\pm$0.010  &  48174.0910$\pm$0.0042  &  0.0010  &  1.52$\pm$0.36 & 0.56$\pm$0.15 &  2.56$\pm$0.42 & \\
$4_4,9/2-3_3,7/2$  &  48175.688$\pm$0.010  &  48175.6827$\pm$0.0045  &  0.0053  &  1.59$\pm$0.19 & 0.51$\pm$0.06 &  2.91$\pm$0.34 & \\

$6_7,15/2-5_6,13/2$&  76626.065$\pm$0.010  &  76626.0628$\pm$0.0035  &  0.0022  & 17.05$\pm$1.07 & 0.44$\pm$0.03 & 36.33$\pm$2.38 & \\
$6_7,13/2-5_6,11/2$&  76629.073$\pm$0.010  &  76629.0901$\pm$0.0032  & -0.0171  & 14.80$\pm$1.08 & 0.51$\pm$0.04 & 27.12$\pm$2.34 & \\

$7_6,11/2-6_5,9/2$ &  79222.404$\pm$0.050  &  79222.4027$\pm$0.0135  &  0.0013  &  2.14$\pm$0.62 & 0.28$\pm$0.14 &  7.29$\pm$1.74 & \\ 
$7_6,13/2-6_5,11/2$&  79226.722$\pm$0.050  &  79226.6778$\pm$0.0134  &  0.0442  &  2.99$\pm$0.85 & 0.61$\pm$0.23 &  4.58$\pm$1.35 & \\ 

$7_7,15/2-6_6,13/2$&  84304.975$\pm$0.015  &  84304.9567$\pm$0.0054  &  0.0183  &  1.79$\pm$0.21 & 0.50$\pm$0.06 &  3.40$\pm$0.44 & \\
$7_7,13/2-6_6,11/2$&  84305.451$\pm$0.015  &  84305.4542$\pm$0.0054  & -0.0032  &  1.16$\pm$0.17 & 0.38$\pm$0.05 &  2.84$\pm$0.43 & \\

$7_8,17/2-6_7,15/2$&  88068.559$\pm$0.010  &  88068.5520$\pm$0.0035  &  0.0070  &  8.16$\pm$1.05 & 0.42$\pm$0.08 & 18.16$\pm$2.28 & \\
$7_8,15/2-6_7,13/2$&  88071.078$\pm$0.010  &  88071.0900$\pm$0.0035  & -0.0120  &  9.56$\pm$0.97 & 0.57$\pm$0.06 & 15.78$\pm$2.18 & \\

$8_7,13/2-7_6,11/2$&  91979.701$\pm$0.050  &  91979.7600$\pm$0.0192  & -0.0590  &                &               &  5.52$\pm$1.46 & B \\ 
$8_7,15/2-7_6,13/2$&  91983.266$\pm$0.030  &  91983.2850$\pm$0.0190  & -0.0190  &  2.32$\pm$0.55 & 0.46$\pm$0.11 &  4.78$\pm$1.40 & \\

$8_8,15/2-7_7,13/2$&  96348.070$\pm$0.020  &  96348.0316$\pm$0.0082  &  0.0384  &  0.54$\pm$0.11 & 0.44$\pm$0.10 &  1.14$\pm$0.28 & \\
$8_8,17/2-7_7,15/2$&  96348.509$\pm$0.020  &  96348.4106$\pm$0.0082  &  0.0984  &  0.89$\pm$0.12 & 0.63$\pm$0.09 &  1.32$\pm$0.26 & \\

$8_9,19/2-7_8,17/2$&  99607.620$\pm$0.010  &  99607.6194$\pm$0.0048  &  0.0006  &  3.60$\pm$0.30 & 0.42$\pm$0.04 &  8.05$\pm$0.84 & \\
$8_9,17/2-7_8,15/2$&  99609.757$\pm$0.010  &  99609.7436$\pm$0.0048  &  0.0134  &  4.99$\pm$0.40 & 0.58$\pm$0.06 &  8.04$\pm$0.84 & \\
\hline
\end{tabular}
\end{center}
\textbf{Notes.}
The observational parameters and their uncertainties were obtained from Gaussian fits to each line profile using \texttt{GILDAS}. \\
\tablefoottext{a}{Observed frequency assuming a systemic velocity of 5.83 km\,s$^{-1}$.}\\
\tablefoottext{b}{Calculated frequency.}\\
\tablefoottext{c}{Integrated line intensity in mK\,km\,s$^{-1}$.}\\
\tablefoottext{d}{Full width at half maximum (FWHM), in km\,s$^{-1}$.}\\
\tablefoottext{e}{Antenna temperature in millikelvin.}\\
\tablefoottext{f}{Fixed value.}\\
\tablefoottext{A}{Affected by a negative feature produced in the folding of the frequency-switching data. The linewidth was fixed to perform the Gaussian fit.}\\
\tablefoottext{B}{Partially blended with the $J_K=5_2-4_2$ transition of CH$_3$CN; only the antenna peak temperature is given.}\\
\end{table*}

\end{appendix}

\end{document}